\documentclass[twocolumn,showpacs,preprintnumbers,amsmath,amssymb,pra]{revtex4}

\usepackage{graphicx}
\usepackage{dcolumn}
\usepackage{bm}

\begin{document}

\title{Scissors mode of a rotating Bose-Einstein condensate}
\author{Marco Cozzini and Sandro Stringari}
\affiliation{Dipartimento di Fisica, Universit\`a di Trento and Istituto
Nazionale per la Fisica della Materia, I-38050 Povo, Italy}
\author{Vincent Bretin, Peter Rosenbusch, and Jean Dalibard}
\affiliation{Laboratoire Kastler Brossel\cite{univ},
24 rue Lhomond, 75005 Paris, France}

\date{\today}

\begin{abstract}
A scissors mode of a rotating Bose-Einstein condensate is investigated both
theoretically and experimentally. The condensate is confined in an
axi-symmetric harmonic trap, superimposed with a small rotating deformation.
For angular velocities larger than $\omega_\perp/\sqrt2\,$, where
$\omega_\perp$ is the radial trap frequency, the frequency of the scissors mode
is predicted to vanish like the square root of the deformation, due to the
tendency of the system to exhibit spontaneous rotational symmetry breaking.
Measurements of the frequency confirm the predictions of theory. Accompanying
characteristic oscillations of the internal shape of the condensate are also
calculated and observed experimentally.
\end{abstract}


\maketitle

Bose-Einstein condensates rotating at high angular velocity exhibit spontaneous
breaking of rotational symmetry \cite{alessio}. This phenomenon, which is the
consequence of two-body repulsive interactions, shows up in the occurrence of
considerable deformations of the trapped atomic cloud in the plane normal to
the rotation axis. These configurations have been recently observed
experimentally using a nearly axi-symmetric harmonic trap, with a small
deformation of the trapping potential rotating around the $z$ axis \cite{ens1}.

Under such conditions the collective modes of the system exhibit
new interesting features. In particular one mode in the transverse
plane (orthogonal to the rotation axis) corresponds to a
shape-preserving oscillation of the atomic cloud with respect to
the principal axes $xy$ of the rotating trap. The frequency
$\omega$ of this scissors-type motion is much smaller than the
mean transverse trap frequency $\omega_\bot$. This surprisingly
low value originates from the fact that the restoring force of the
oscillation is proportional to the small trap deformation, while
the moment of inertia of the condensate remains finite due to its
considerable deformation. This represents a major difference with
respect to the traditional scissors mode \cite{david,Marago} of a
non-rotating condensate where both the restoring force and the
moment of inertia vanish in the limit of an axi-symmetric trap.

The purpose of this work is to provide both a theoretic and
experimental investigation of the problem. We first derive the
relevant equations describing the scissors mode. For a rotation
frequency $\Omega$ larger than the critical value
$\omega_\bot/\sqrt2\,$, we show that the scissors mode frequency
$\omega$ vanishes like the square root of the trap deformation. We
then report on the experimental observation of this scissors mode,
and we present results which confirm the theoretic predictions
with good accuracy.

The atoms of mass $m$ are confined in the transverse plane by the potential
\begin{equation}
V_\bot(x,y)=m(\omega_x^2x^2+\omega_y^2y^2)/2\ , \label{potential}
\end{equation}
where $ \omega_{x,y}^2=\omega_\bot^2(1\pm\epsilon)\;$. The $(x,y)$
coordinates in the rotating frame are deduced from the coordinates
in the laboratory frame by a rotation $\Omega t$ and the trap
deformation $\epsilon$ is positive.  A simple description of the
collective oscillations of a rotating condensate is provided by
the hydrodynamic equations of superfluids evaluated in the
rotating frame:
\begin{equation} \label{eq:HD dens rot}
\frac{\partial\rho}{\partial t}+\mbox{\boldmath$\nabla$}
\cdot[\rho({\bf v}-{\bf \Omega}\wedge{\bf r})]=0\ ,
\end{equation}
\begin{equation} \label{eq:HD vel rot}
m\frac{\partial{\bf v}}{\partial t}+\mbox{\boldmath$\nabla$}
\left[\frac{mv^2}{2}+V+ g\rho-m{\bf v}
\cdot({\bf \Omega}\wedge{\bf r})\right]=0\ ,
\end{equation}
where $\rho({\bf r},t)$ is the spatial density, ${\bf v}({\bf r},t)$ is the
velocity field in the laboratory frame, and ${\bf \Omega}=\Omega\,{\bf u}_z$
(${\bf u}_z$ unit vector along the $z$ axis). The harmonic confining potential
$V({\bf r})=V_\bot(x,y) + m\omega_z^2z^2/2$ is time independent in the rotating
frame. The parameter $g$ characterizes the strength of the interatomic
interactions and is related to the $s$-wave scattering length $a$ by
$g=4\pi\hbar^2a/m\,$. These equations are valid in the Thomas-Fermi limit,
where the so-called quantum pressure term can be neglected in (\ref{eq:HD vel
rot}) \cite{review}. We shall be interested here in vortex-free solutions for
which $\mbox{\boldmath$\nabla$}\wedge{\bf v}=0\,$.

\begin{figure}[t]
\begin{center}
\includegraphics[width=8.5cm]{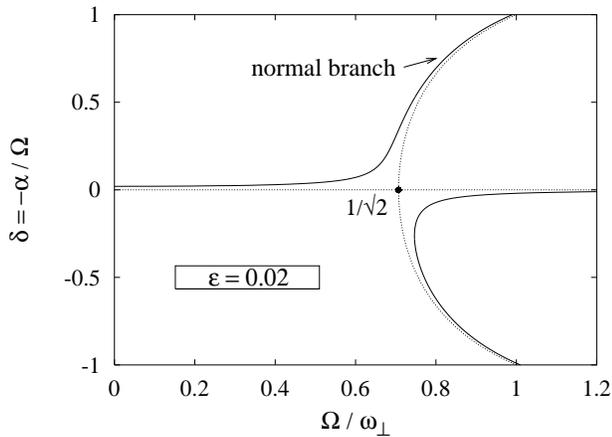}
\end{center}
\caption{Deformation $\delta$ of the cloud for the steady-state
solutions of Eqs. (\ref{eq:HD dens rot}) and (\ref{eq:HD vel rot})
as a function of the angular velocity $\Omega$ of the trap for a
fixed value of the trap deformation $\epsilon\,$. The dotted line
represents the solution for $\epsilon=0\,$.} \label{fig:branches}
\end{figure}

Stationary solutions of these equations can be obtained in the form ${\bf
v}_0({\bf r}) = \alpha \mbox{\boldmath$\nabla$} (xy)$ for the velocity field
and $\rho_0({\bf r})=\left[\tilde\mu-m
\left(\tilde\omega_x^2x^2+\tilde\omega_y^2y^2 +
\omega_z^2z^2\right)/2\right]/g$ for the density profile. The density profile
has the form of an inverted parabola whose parameters ($\tilde\mu\,$,
$\tilde\omega_x\,$, $\tilde\omega_y$) are determined in a self-consistent way
as a function of $\Omega$ \cite{alessio}. The value of the parameter $\alpha$
is determined by the solution(s) of the cubic equation $\alpha^3 +
\alpha(\omega_\perp^2-2\,\Omega^2) + \epsilon\omega_\perp^2\Omega = 0\,$. It is
related to the deformation $\delta=\langle y^2-x^2 \rangle/\langle x^2+y^2
\rangle$ of the atomic cloud by the simple expression $\alpha=-\Omega\delta\,$.
In the present work we will be interested in the so-called normal branch
\cite{alessio}, corresponding to the stationary solutions which can be obtained
by an adiabatic increase of the angular velocity of the trap, starting from
$\Omega=0$ (see Fig.~\ref{fig:branches}). For $\Omega \ge \omega_\perp/\sqrt2$
these solutions exhibit large values of the cloud deformation ($\delta \sim 1$)
even if the trap deformation $\epsilon$ is much smaller than 1.

An important class of collective oscillations can be derived on top of such
stationary solutions by looking for time dependent solutions of the form
$\delta\rho({\bf r}) = a_0+a_xx^2+a_yy^2+a_zz^2+a_{xy}xy$ and $\delta{\bf v}
({\bf r}) =
\mbox{\boldmath$\nabla$}(\alpha_xx^2+\alpha_yy^2+\alpha_zz^2+\alpha_{xy}xy)$
where $a_i$ and $\alpha_i$ are time dependent parameters to be determined by
solving  Eqs. (\ref{eq:HD dens rot}) and (\ref{eq:HD vel rot}). In the linear
limit, one can look for solutions varying in time like $e^{-i\omega t}$. The
collective frequencies $\omega$ are then found to obey the equation
\begin{equation}
  \label{eq:det(A)=0}
  \omega^8+c_3\omega^6+c_2\omega^4+c_1\omega^2+c_0 = 0
\end{equation}
with the coefficients $c_i$  given in the appendix. In \cite{alessio} it has
been shown that these collective oscillations are dynamically stable when
evaluated on the normal branch. The stability conditions for higher multipole
oscillations have been studied in \cite{sinha}, where it has been shown that
the normal branch becomes dynamically unstable against the production of such
excitations for $\Omega \gtrsim 0.8\,\omega_\perp\,$ when $\epsilon \to 0\,$.

For the two external branches (see Fig.~\ref{fig:branches}), one finds that the
coefficient $c_0$ in (\ref{eq:det(A)=0}) vanishes linearly with $\epsilon$ as
$\epsilon \to 0\,$. In this limit, one of the eigenfrequencies of
(\ref{eq:det(A)=0}) tends to $0$, while the others remain finite. The frequency
$\omega$ of the corresponding ``soft" mode is:
\begin{equation} \label{eq:omega}
\omega^2 = -\,\frac{c_0}{c_1} =
\pm\frac{10\,\omega_\perp^2\Omega(\omega_\perp^2-\Omega^2)
\sqrt{2\,\Omega^2-\omega_\perp^2}}
{3\,\omega_\perp^4+3\,\omega_\perp^2\Omega^2+2\,\Omega^4}\,\epsilon\ .
\end{equation}
The positive value for $\omega^2$ corresponds to the normal branch while the
negative value, which is a signature of dynamical instability, corresponds to
the opposite branch. For the normal branch, the frequency $\omega$ of the low
energy mode vanishes like the square root of the trap deformation $\epsilon\,$,
as announced in the introduction of this paper. It is worth noticing that this
frequency does not depend on the value of $\omega_z\,$. By looking at the
corresponding form of the solution for $\delta \rho$ and $\delta {\bf v}\,$,
one can show that it corresponds to a scissors mode, i.e. to the oscillation of
the angle $\theta$ between the axes of the condensate and the principal axes
$x,y$ of the trap. The oscillation of $\theta$ is accompanied by an oscillation
of the condensate deformation $\delta$ around the equilibrium value $\delta_0$.
This oscillation of $\delta$ is dephased by $\pi/2$ with respect to the one of
the angle $\theta\,$. For small trap deformations we find
\begin{equation} \label{eq:delta}
\delta(t)-\delta_0=\frac{\omega_\perp^2\,\dot\theta}
{\Omega^2\,\sqrt{2\,\Omega^2-\omega_\perp^2}}\ .
\end{equation}
\indent The fact that the frequency of the scissors mode vanishes when
$\epsilon=0$ (see Eq.~(\ref{eq:omega})) reflects the rotational invariance of
the Hamiltonian. This behavior deeply differs from the scissors mode in a non
rotating condensate, whose frequency is given by $\omega_\bot \sqrt{2}$ and
does not vanish in the limit of a symmetric trap \cite{david}. The result
(\ref{eq:omega}) holds for a small amplitude $\Delta \theta$ of the oscillatory
motion of $\theta\,$. To study motions with a larger amplitude, one has to
solve numerically the hydrodynamic equations (\ref{eq:HD dens rot}) and
(\ref{eq:HD vel rot}). We find that the frequency of the motion decreases as
$\Delta \theta$ increases, as for a simple gravitational pendulum.

We now turn to the experimental observation of this scissors mode.
We use a $^{87}$Rb gas in a Ioffe-Pritchard magnetic trap, with
frequencies $\omega_x /2\pi = 182$~Hz, $\omega_y \simeq \omega_x$,
and $\omega_z/2\pi = 11.7$~Hz. The cloud is pre-cooled using
optical molasses to a temperature $\sim\;100\;\mu$K. The gas is
further cooled by radio-frequency evaporation to a temperature
around $50$~nK, corresponding to a quasi-pure condensate with
$10^5$ atoms. We denote by $t_0$ the time at which the evaporation
phase ends. For time $t>t_0$, the atomic cloud is stirred by a
focused laser beam of wavelength $852$~nm and waist $w_0=20 \mu$m,
whose position is controlled using acousto-optic modulators
\cite{ens1}. This laser beam creates a rotating optical-dipole
potential which is harmonic over the extension of the cloud. The
superposition of this dipole potential and the magnetic one
creates a transverse trapping potential identical to
(\ref{potential}). The trap deformation $\epsilon$ is proportional
to the laser intensity $I_L$ and can be adjusted between 0 and
$4\;\%$.

The displacement of the center of the trap due to gravity and
slight asymmetries in the trapping geometry produce an additional
static deformation which has been estimated to be $\sim1\%$ by
measuring the splitting between the center of mass frequencies
along the $x$ or $y$ axes. This static anisotropy plays a minor
role in the present study and has been neglected in the analysis
above.

The frequency $\Omega(t)$ of the stirrer is first varied linearly
during the time interval $(t_0,t_1)$, starting from
$\Omega(t_0)=0$ up to $\Omega(t_1)=2\pi\times 139$~Hz (so that
$\Omega(t_1)\sim 0.76\, \omega_\bot$). The stirring frequency then
stays constant in the time interval $(t_1,t_2)$. At time $t_2$ we
switch off the magnetic trap and the laser stirrer, allow for a
25~ms free-fall, and image the absorption of a resonant laser beam
propagating along $z$. We measure in this way the transverse
density profile of the atom cloud, which we fit assuming a
parabolic shape. We extract from the fit the long and short
diameters in the transverse plane (hence the deformation
$\delta(t)$), and the orientation $\theta(t)$ of the condensate
axes with respect to the rotating frame of the laser stirrer.

The excitation of the scissors mode arises directly from the small
non-adiabatic character of the condensate evolution as $\Omega(t)$
increases, during the time interval $(t_0,t_1)$. At time $t_1$,
the state of the condensate slightly differs from the steady-state
corresponding to $\Omega(t_1)$. Consequently the state of the
condensate still evolves in the rotating frame in the time
interval $(t_1,t_2)$, even though the characteristics of the
stirrer do not change anymore.

\begin{figure}[t]
  \includegraphics[width=8.5cm]{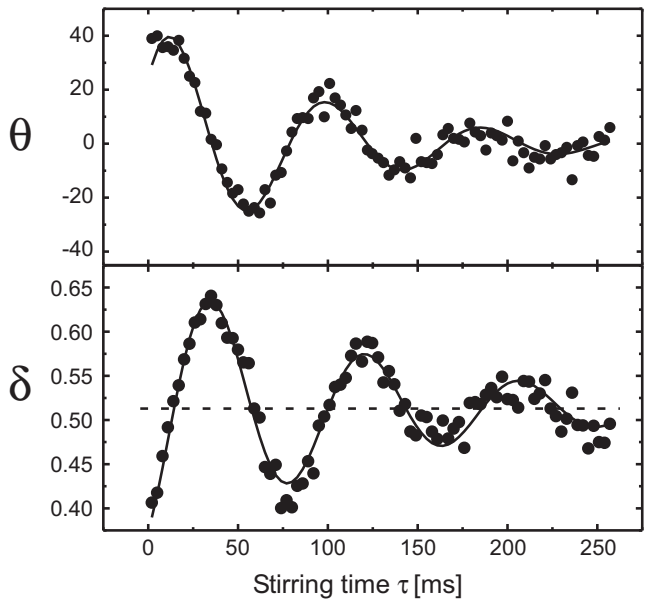}
\caption{Variation of the angle $\theta$ (in degrees) and of the
condensate deformation $\delta$ with the stirring time
$\tau=t_2-t_1$. The oscillations of $\theta$ and $\delta$ are a
signature of the scissors mode. The oscillation frequency $\omega/
2\pi\simeq 11.6$~Hz is much smaller than the transverse trap
frequency $\omega_\bot/2\pi$, in agreement with
Eq.~\ref{eq:omega}. The solid lines are a fit to the data with a
damped sine function. The dashed line in the lower graph shows the
stationary value $\delta=\delta_0$.}
  \label{fig:evolution}
\end{figure}

We have plotted in Fig.~\ref{fig:evolution} the angle $\theta$ and
the condensate deformation $\delta$ as a function of the time
$\tau=t_2-t_1$. These data have been obtained for a ramping time
$t_1-t_0=360$~ms and a spoon anisotropy $\epsilon=0.017\,(6)$. The
angle $\theta$ oscillates with a frequency $\omega/2 \pi=
11.6\;(\pm 0.1)$~Hz, in good agreement with the value $11.4\;(\pm
2.1)$~Hz expected from (\ref{eq:omega}). The initial amplitude
$\Delta \theta$ is 43 degrees, and the oscillation is damped with
a time constant $\sim 110$~ms. The deformation $\delta$ of the
condensate also exhibits a small oscillatory motion with an
amplitude $\Delta \delta=0.16$ around the mean value
$\delta_0=0.51$. This motion has the same frequency as that of
$\theta$, is phase shifted by $\sim \pi/2$, as expected from
(\ref{eq:delta}) for the scissors mode, and it is damped with a
similar time constant.

\begin{figure}
  \includegraphics[width=8.5cm]{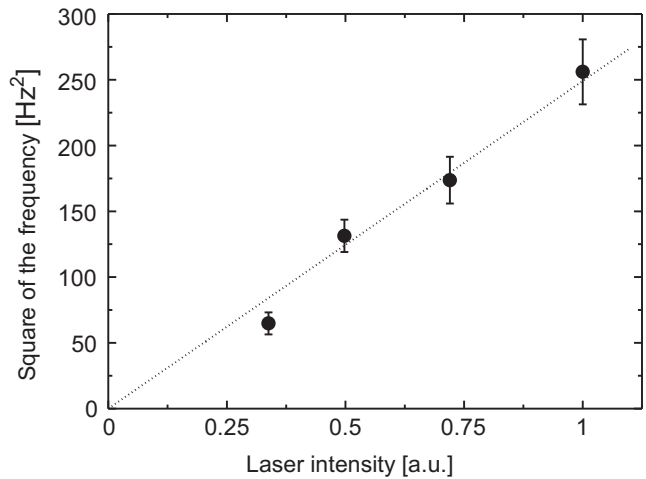}
  \caption{Variation of the square of the scissors mode frequency $\omega/2\pi$
  with the laser intensity $I_L$ (arbitrary units). The error bars indicate the
  statistical spread of the results obtained for various runs of the experiment.
  The dotted line is a linear
  fit according to the prediction of Eq.~\ref{eq:omega}.}
  \label{fig:intensity}
\end{figure}

To confirm that the oscillatory motion shown in
Fig.~\ref{fig:evolution} indeed corresponds to the scissors mode
described above, we have measured the variation of the frequency
$\omega$ as a function of the laser intensity, which is itself
proportional to the trap anisotropy $\epsilon\,$. The data are
plotted in Fig.~\ref{fig:intensity}. They clearly show the
expected dependance $\omega^2\propto\epsilon\,$. Actually, this
scissors mode constitutes a very precise way to measure
$\epsilon\,$, once the frequencies $\omega_\perp$ and $\Omega$ are
known with sufficient precision.

We have also compared the measured amplitudes $\Delta \theta(t_2)$
and $\Delta \delta(t_2)$ with the results from a numerical
integration of the equations of motion for the coefficients $a_i$
and $\alpha_i$ when $\Omega$ is ramped from 0 up to its final
value $\Omega(t_1)$. The calculated results reproduce the behavior
found experimentally. The amplitude of the scissors mode decreases
when the ramping time $t_1-t_0$ or the trap anisotropy $\epsilon$
are increased. The physical reason for this variation is clear:
the evolution of the condensate during the time interval
$(t_0,t_1)$ is closer to adiabatic following, and the condensate
is left at time $t_1$ in a state closer to the stationary state
expected for $\Omega=\Omega(t_1)$. However we could not reach a
strict quantitative agreement between the calculated amplitudes
and the measured ones (typical deviation of 50\%). We think that
this is due to the damping of the scissors mode, present in the
experiment (see Fig.~\ref{fig:evolution}) and neglected in our
simple theoretical model. A proper description of this damping
could be obtained using a formalism similar to \cite{Zaremba},
where the damping of the scissors mode in a static trap is
investigated.

\begin{figure}[t]
\includegraphics[width=8.5cm]{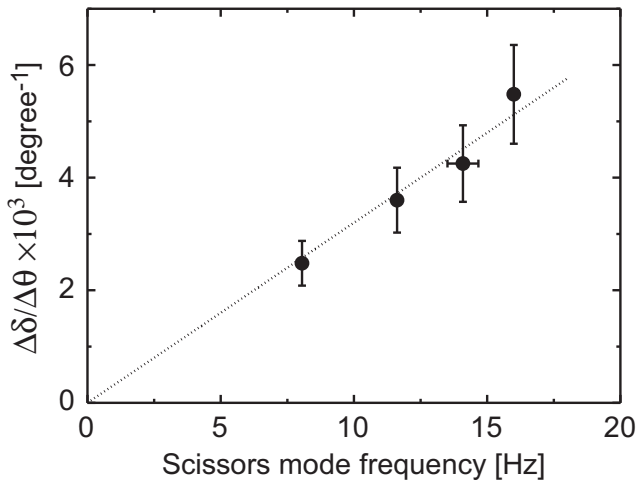}
\caption{Variation of the ratio $\Delta \delta/\Delta \theta$ as a
function of $\omega/2\pi\,$. The point with an horizontal error
bar corresponds to the average of three results with very similar
scissors frequencies $\omega$. For small $\omega\,$, corresponding
to $\epsilon\to0$, the oscillation of the cloud deformation is
negligible and the condensate motion is similar to that of a rigid
body. The dotted line is a linear fit to the data.}
\label{fig:amplitude}
\end{figure}

Our final study concerns the behavior of the ratio
$\Delta\delta/\Delta\theta$ at $t_2$, where $\Delta\delta$ and
$\Delta\theta$ are the oscillation amplitudes for the deformation
and for the angle respectively.  The results, plotted in
Fig.~\ref{fig:amplitude}, show that $\Delta\delta/\Delta\theta$
varies linearly with $\omega/2\pi\,$ with a slope of $3.2\times
10^{-4}$ [degree.Hz]$^{-1}$ (i.e. $\Delta\delta/(\omega \Delta
\theta) \simeq 2.9\times 10^{-3}$~s/radian$^2$). This is in good
agreement with the prediction of Eq.~(\ref{eq:delta}) which gives
an expected slope of $4.0\times 10^{-4}$ [degree.Hz]$^{-1}$. The
small deviation between predicted and experimental slopes may be
due to nonlinearities originating from the relatively large
amplitude of the excitation. This linear dependance of
$\Delta\delta/\Delta\theta$ on the scissors frequency confirms the
idea that for almost  symmetric traps, where $\epsilon\to0$ and
thus $\omega\to0\,$, the motion of the condensate is
shape-preserving, similar to that of a rigid body.

To summarize, in this paper we have presented the theoretic analysis and the
experimental observation of a new mode of a trapped rotating condensate. The
existence of this mode relies upon the breaking of the rotational symmetry of
the condensate caused by atom interactions. The rotating condensate can have a
sizeable deformation even in the limit of an arbitrarily small stirring
potential, and this gives unique properties to the mode of interest, such as a
frequency much smaller than the trap frequency. Here we have been mostly
interested in the small oscillations of the condensate around its rotating
steady-state. A natural extension of this work consists in studying the
non-linear regime of the system, where a chaotic dynamics can emerge.

\begin{acknowledgments}
P. R. acknowledges support by the Alexander von Humboldt-Stiftung and by the
EU, contract no. HPMF-CT-2000-00830. S. S. likes to thank the hospitality of
the Laboratoire Kastler Brossel.

This work was partially supported by the R\'egion Ile de France, CNRS,
Coll\`{e}ge de France, DRED and the EU, contract no. HPRN-CT-2000-00125 and by
the Ministero dell'Universit\`a e della Ricerca Scientifica e Tecnologica
(MURST).
\end{acknowledgments}

\vskip 1cm
\appendix{\large \bf \centerline {Appendix}} \vskip 3mm
The values of the coefficients $c_i$ entering into (\ref{eq:det(A)=0}) are:
\begin{eqnarray*}
c_3 & = & -\omega_\perp^2\{8(1+\tilde\Omega^2)+3\,\lambda^2\} \ ,\\
c_2 & = & \omega_\perp^4\{4(5-2\,\epsilon^2)+16\,\tilde\Omega^2+16\,\tilde\Omega^4\\
&&-4[\tilde\alpha^2(2\,\tilde\Omega^2-1)-3\,\tilde\alpha\tilde\Omega\epsilon]\\
&&+2\,\lambda^2(11-\tilde\alpha^2+12\,\tilde\Omega^2)\} \ ,\\
c_1 & = & -\omega_\perp^6\{16[(2\,\tilde\Omega^2-1)(2\,\tilde\Omega^2-1+\epsilon^2)\\
&&-\tilde\alpha^2(2\,\tilde\Omega^2-1+3\,\epsilon^2)-\tilde\alpha\tilde\Omega\epsilon
(5-4\,\tilde\Omega^2)-3\,\tilde\Omega^2\epsilon^2]\\
&& +4\,\lambda^2[(13-5\,\epsilon^2)+8\,\tilde\Omega^2+12\,\tilde\Omega^4\\
&& -\tilde\alpha^2(2\,\tilde\Omega^2-1)+3\,\tilde\alpha\tilde\Omega\epsilon]\}\ , \\
c_0 & = & \omega_\perp^8\{40\,\lambda^2[(2\,\tilde\Omega^2-1)(2\,\tilde\Omega^2-1+\epsilon^2)\\
&& -\tilde\alpha^2(2\,\tilde\Omega^2-1+3\,\epsilon^2)-\tilde\alpha\tilde\Omega\epsilon(5-4\,\tilde\Omega^2)-
3\,\tilde\Omega^2\epsilon^2]\} \ ,
\end{eqnarray*}
where we have introduced  the reduced quantities
$\tilde\Omega=\Omega/\omega_\perp\,$, $\lambda=\omega_z/\omega_\perp\,$ and
$\tilde\alpha=\alpha/\omega_\perp\,$.



\begin{thebibliography}{99}

\bibitem[*]{univ} Unit\'e de Recherche de l'Ecole normale sup\'erieure et
de l'Universit\'e Pierre et
Marie Curie, associ\'ee au CNRS

\bibitem{alessio}
A.~Recati, F.~Zambelli, and S.~Stringari, Phys. Rev. Lett.
{\bf86}, 377 (2001).

\bibitem{ens1}
K.~W.~Madison, F. Chevy, V. Bretin, and J. Dalibard, Phys. Rev.
Lett. {\bf 86}, 4443 (2001).

\bibitem{david}
D.~Gu\'ery-Odelin and S.~Stringari, Phys. Rev. Lett. {\bf 83},
4452 (1999).

\bibitem{Marago}
O. M. Marag\`o, S. A. Hopkins, J. Arlt, E. Hodby,
G.~Hechenblaikner, and C. J. Foot, Phys. Rev. Lett. {\bf 84}, 2056
(2000).

\bibitem{review}
F. Dalfovo, S. Giorgini, L. Pitaevski, and S. Stringari, Rev. of
Mod. Phys. \textbf{71}, 463 (1999).

\bibitem{sinha}
S.~Sinha and Y.~Castin, Phys. Rev. Lett. {\bf87}, 190402 (2001).

\bibitem{Zaremba}
B. Jackson and E. Zaremba, Phys. Rev. Lett. \textbf{87}, 100404
(2001).

\end{thebibliography}
\end{document}